\documentclass[aip,jcp,reprint,numerical,superscriptaddress]{revtex4-1}
\usepackage[caption=false]{subfig}
\usepackage{pslatex}
\usepackage{graphicx}
\usepackage{verbatim}
\usepackage[breaklinks=true,colorlinks,citecolor=blue, linkcolor=blue,urlcolor=blue]{hyperref}
\usepackage{amsmath,amssymb}
\usepackage{booktabs}
\usepackage{physics}
\usepackage{csquotes}
\newcommand{\vect}[1]{\textbf{#1}}
\usepackage[T1]{fontenc}
\usepackage{ae,aecompl}
\usepackage{times}
\begin{document}
\title{ Accurate effective potential for density amplitude and the corresponding Kohn-Sham exchange-correlation potential calculated from approximate wavefunctions }
\author{Ashish Kumar}
\email[]{ashishkr@iitk.ac.in}
\affiliation{Department of Physics, Indian Institute of Technology Kanpur, Kanpur-208016, India}   
\author{Rabeet Singh}
\email[]{rabeet@niser.ac.in}
\affiliation{School of Physical Sciences, National Institute of Science Education and Research, HBNI, Bhubaneswar 752050, India}     
\author{Manoj K. Harbola}
\email[]{mkh@iitk.ac.in}
\affiliation{Department of Physics, Indian Institute of Technology Kanpur, Kanpur-208016, India}
\date{\today}
\begin{abstract}
Over the past few years it has been pointed out that  direct inversion of accurate but approximate ground state densities leads to Kohn-Sham exchange-correlation (xc) potentials that can differ significantly from the exact xc potential of a given system. On the other hand, the corresponding wavefunction based construction of exchange-correlation potential as done by Baerends et al. and Staroverov et al. obviates  such problems and leads to potentials that are very close to the true xc  potential. In this paper, we provide an understanding of why the wavefunction based approach  gives the  exchange-correlation potential accurately. Our understanding is  based on the work of Levy, Perdew and Sahni (LPS) who gave an equation for the square root of density (density amplitude) and the expression for the associated effective potential in the terms of the corresponding wavefunction. We show that even with the use of approximate wavefunctions the LPS expression gives accurate effective and exchange-correlation potentials. Based on this we also identify the source of difference between the potentials obtained from a wavefunction and those given by the inversion of the associated density. Finally, we suggest  exploring the possibility of obtaining accurate ground-state density from an approximate wavefunction for a system by making use of the LPS effective potential.      		
\end{abstract}

\maketitle
\section{Introduction}
Density functional theory (DFT) \cite{Hohn,Kohn_1965} is the most widely used theory  \cite{burke_1_An_rev_2015} of electronic structure and is applied to study systems of all sizes, from atoms to bulk solids.   Application of the theory, however, requires  approximating the exchange-correlation energy functional and it is usually assumed  that better and better approximations for this energy functional will also lead to more accuracy for  other quantities. On the other hand, it has been noted \cite{Medv_2017} that densities do not necessarily improve with improvement in the energy. In light of such observations and the fact that the theory can be applied only approximately (albeit yielding accurate answers with better functionals), it is imperative that exact results be obtained wherever possible. This helps in gaining \cite{Buijse_1989,Gritsenko_1996,Teal2,Teal3,Teal4,Makmal_2011,Wagner_2012,2014_Gould,Kohut_2016,Proetto_2016,Godby_PRA_2016,Rabeet_2017,Staroverov_PNAS_18,kummel_18,2019_Gould} insights into how the theory works in different situations. As such many studies  \cite{Stott_1988,Gorling_1992,Zhao_1992, Zhao_1993,Wang_1993, Zhao_1994,Vlb_1994, Umrigar_PRA_1994,Vlb_1995,handy_1996,Ingamells_CPL.248.373,Mura_jcp_1997,WY1, WY2,Peirs_2003,Stott_2004, Viktor_2012, Viktor_2013, Wagner_2014, Viktor_2015, Hollins_2017,Wasserman_2017} have been carried out that obtain the  Kohn-Sham potential for a given near-exact density for a variety of  many-electron systems. These densities are obtained by solving the many-body Schrödinger equation as accurately as possible by many different methods, such as integration of the Schrödinger equation directly  or  application of the variational method. The latter uses the variational principle  with approximately chosen parameterized wavefunction \cite{Hylleraas_ZP.48.469, Hylleraas_ZP.54.347, Hylleraas_ZP.65.209, Kinoshita_PR.105.1490, Kinoshita_PR.115.366,Koga_JCP.96.1276, Koga_IJQC.46.689, Frankowski_PR.146.46, Fruend_PRA.29.980,Umrigar_PRA.50.3827,Chandrasekhar_AJ.100.176, Sech_JPB.30.L47,Sech_PRA.63.022501,Chauhan_CPL.639.248,Tripathy_JPB.28.L41,Wu_PRA.26.1762,Bhattacharyya_JPB.29.L147,2005_Sarsa_JCP,2005b_Sarsa_JCP,zen_2013,Kim_2018}. The method of choice in applying the variational scheme, however, is expanding the wavefunction in terms of a basis set and optimizing  the expansion parameters. From this wavefunction the density of the system is obtained.  Before we proceed further from this point to present our work, we go over some definitions that are going to be used in the paper.
\par The exact wavefuncion of a system of N interacting electron in an external potential $v_{ext}(\vect{r})$ is obtained by solving the time-independent Schrödinger equation
\begin{equation}
H_N \Psi =E \Psi \label{I11a}
\end{equation}
for the wavefunction $\Psi(\vect{x}_{1-N})$ where $\vect{x} = (\vect{r},\sigma) $ denotes the space $(\vect{r})$, spin  $(\sigma)$ variables electron respectively  and $\vect{x}_{i-j} = \vect{x}_{i},\vect{x}_{i+1} \cdots \vect{x}_{j}$. Here (atomic-units are used throughout the paper)
\begin{equation}
H_N = \sum_{i=1}^{N} \Big( -\frac{1}{2}\nabla^2_i +v_{ext}(\vect{r}_i) \Big) +\frac{1}{2} \sum \limits_{ \underset{i \ne j}{i,j =1} }^{N} \frac{1}{|\vect{r}_i- \vect{r}_j|} \label{I11b}
\end{equation}
is the Hamiltonian and the eigenvalue $E$ gives the energy of the system. The density $\rho(\vect{r})$ corresponding to the wavefunction $\Psi(\vect{x}_{1-N})$ is given by
\begin{equation}
\rho(\vect{r)}) = \int |\Psi (\vect{x},\vect{x}_{2-N})|^2 d \sigma d \vect{x}_{2-N}. \label{I12}
\end{equation}
Now according to the Hohenberg-Kohn theorem \cite{Hohn} there is a one-to-one correspondence between the external potential $v_{ext}(\vect{r})$ and the ground state density $\rho_0(\vect{r})$ of a system obtained from ground state wavefunction $\Psi_0(\vect{x}_{1-N})$ by using Eq. (\ref{I12}). Thus either $v_{ext}(\vect{r})$ or $\rho_0(\vect{r})$  can be used to specify a system. The ground state  density for a system can also be obtained by solving self-consistently the Kohn-Sham equation \cite{Kohn_1965}
\begin{equation}
[- \frac{1}{2} \nabla^2 + v_{ext}(\vect{r}) + v_{H}(\vect{r}) +v_{xc}(\vect{r}) ] \phi_i(\vect{r}) = \epsilon_i \phi_i(\vect{r})  \label{I13}
\end{equation}
where 
\begin{equation}
v_H(\vect{r}) = \int \frac{\rho(\vect{r}')}{|\vect{r}' -\vect{r}|}d\vect{r}'  \label{I14}
\end{equation}
is the Hartree potential for a density $\rho(\vect{r})$ and 
\begin{equation}
v_{xc}(\vect{r})  = \frac{\delta E_{xc}[\rho]}{\delta\rho(\vect{r})} \label{I15}
\end{equation}
is exchange-correlation potential, calculated as the functional derivative of the exchange-correlation energy functional $ E_{xc}[\rho]$. The self consistent solution of the Kohn-Sham equation gives the orbitals $\{ \phi_i^0 (\vect{r})\}$ that leads to the density through the formula $\rho_0(\vect{r}) = \sum_{i} |\phi_i^0 (\vect{r})|^2 $. However, since functional $ E_{xc}[\rho]$ is not known, the corresponding exchange-correlation potential for a given ground state density can not be calculated exactly using Eq. (\ref{I15}). Thus other techniques have to be developed to get this potential for a given system. In the following we will denote exact exchange-correlation potential for a given external potential alternatively as $v_{xc}[v_{ext}](\vect{r})$ or $v_{xc}[\rho_{0}](\vect{r})$, where it is understood that $\rho_0(\vect{r})$ is the ground-sate  density corresponding to $v_{ext}(\vect{r})$.
\par  To calculate the exact Kohn-Sham exchange-correlation potential $v_{xc}(\vect{r})$ from a given density $\rho(\vect{r})$, the most straightforward method would be to  invert the density  numerically.
In the following we will denote the exchange-correlation potential obtained by inversion of density as $v_{xc}^{\rho}(\vect{r})$. There are several methods \cite{Werden, Stott_1988,Gorling_1992, Zhao_1992,Wang_1993, Zhao_1993, Zhao_1994, Wang_1993, Vlb_1994, Schipper1997, WY2,Peirs_2003, Stott_2004, Wagner_2014,Hollins_2017,Wasserman_2017,Finzel2018}  proposed for this  inversion  and most of them have been shown \cite{Kumar_2019} to emanate from a single algorithm based on the Levy-Perdew-Sahni (LPS) equation \cite{LPS_1984} for the square root of the density. However, these methods are highly sensitive to the correctness of  the density and its derivatives for a given $v_{ext}(\vect{r})$ and can lead to $v_{xc}^{\rho}(\vect{r})$ having spurious features in them \cite{Schipper1997,Mura_jcp_1997,savin_2003,2011_Jacob_jcp,2013_jacob_jcp,Staroverov_2017b}. It is easy to understand why this happens: inversion algorithms give the exact potential corresponding to a density and hence even an extremely  small deviation of density $\rho(\vect{r})$ from the exact one $\rho_0(\vect{r})$ could lead to very different potentials \cite{savin_2003}. For example, when densities obtained from wavefunctions expressed in Gaussian basis sets are used, one observes \cite{Schipper1997,Mura_jcp_1997,Staroverov_2017b} large oscillations in the exchange-correlation potentials of atom near the nucleus and the potential increases indefinitely in the asymptotic region. This is despite the corresponding density $\rho(\vect{r})$ being close to the exact density $\rho_0(\vect{r})$. In
this connection, we note it has also been pointed out \cite{Gidopoulos_2012} that use of
finite basis in construction of the optimised effective potential (OEP) leads to oscillations in the resulting potential, although for
entirely different reasons. Consequently, even with extended basis like
plane-waves, very large basis sets containing several thousands of plane
waves have to be used for carrying out the calculation of OEP. On the other hand an alternate approach that has been proposed \cite{Gritsenko_1998,Schipper_1998,Viktor_2013, Viktor_2015} is to use the wavefunctions directly to get the Kohn-Sham potential. So far all applications of this approach \cite{Viktor_2013,Viktor_2015,Viktor_2015_JCP,Staroverov_2017b,Ospadov_2017,Ospadov_2018,Staroverov_PNAS_18} have shown that  for a given nearly exact wavefunction $\Psi$, it leads to the exchange-correlation potential-we denote it as $v_{xc}^{\Psi}(\vect{r})$- that is very close to the exact potential $v_{xc}[\rho_0](\vect{r})$ and is free of  undesirable features that appear when the corresponding  density $\rho(\vect{r})$ is inverted. This has been attributed to the potential in wavefunction approach being the \enquote{sum of commensurate,
well-behaved terms} \cite{Viktor_2013} by Staroverov et al. However a perspicuous understanding of why these terms are well behaved is missing. 
\par The purpose of the present paper is to provide an insight into why the wavefunction based method works better even-with wavefunctions which are close to but not exact and how the potential $v_{xc}^{\Psi}(\vect{r})$  obtained through it is connected to the true exchange-correlation potential of a system. For this we make use of the compact expression given by Levy-Perdew $\&$ Sahni (LPS) \cite{LPS_1984} and other researchers \cite{march_1985_b,1987_March,1986_Hunter,Buijse_1989,1989_Deb_Chattaraj,Gritsenko_1994} for the effective potential for the square-root of density and reach our conclusions based on this formula.  We focus on the LPS expression because  the wavefunction based formulae \cite{Buijse_1989} for the exchange-correlation potential  given in different forms  are ultimately related to this expression. Therefore in the following we start our presentation with a discussion of the LPS equation and then derive our main result based on it.   
\section{The LPS equation and its application}  
\subsection{The LPS equation and the corresponding potential $v_{eff}(\vect{r})$}
The LPS equation  \cite{LPS_1984} satisfied by the square root of  the ground-state density $\rho_0(\vect{r})$ for  $N$ electrons corresponding to the Hamiltonian of Eq. (\ref{I11b}) is 
\begin{equation}
 \Big [-\frac{1}{2} \nabla^2+v_{ext}(\vect{r})+v_{eff}(\vect{r}) \Big ]\rho_0^{1/2}(\vect{r}) = \mu\rho_0^{1/2}(\vect{r}) \label{scheq_den}
\end{equation}
where $\mu$ is the chemical potential and  effective potential $v_{eff}(\vect{r})$ is calculated from the wavefunction $\Psi_0$ as
\begin{equation}
\begin{split} 
v_{eff}[\Psi_0](\vect{r})& = \int  \frac{\rho_{N-1}(\vect{r};\vect{r}')}{|\vect{r}-\vect{r}'|} d \vect{r}' \\ & \quad +\langle \Phi^0_{N-1}|H_{N-1}-E^0_{N-1} |\Phi^0_{N-1}\rangle \\ 
 & \quad \quad +\frac{1}{2}\int |\nabla \Phi^0_{N-1} |^2d\sigma d \vect{x}_{2-N}. \label{LPS_mb}
\end{split}
\end{equation}
Here $H_{N-1}$ is Hamiltonian of $N-1$ interacting electrons moving in external potential potential $v_{ext}(\vect{r})$ and $E^0_{N-1}$ is corresponding ground- state energy.
 $\rho_{N-1}(\vect{r};\vect{r}')$ is density of $N-1$ particles at $\vect{r}'$ associated with the function  $\Phi^0_{N-1}(\vect{x},\vect{x}_{2-N})$. Thus  
\begin{equation}
\begin{split}  
\rho_{N-1}(\vect{r};\vect{r}') & = \\&(N-1) \int |\Phi^0_{N-1}(\vect{x},\vect{x}',\vect{x}_{3-N}) |^2 d\sigma d\sigma' d\vect{x}_{3-N}.
\end{split}
\end{equation}
Here the function
\begin{equation}
 \Phi^0_{N-1}(\vect{x},\vect{x}_{2-N})= \Big( \frac{N}{\rho_0(\vect{r})}\Big)^{1/2} \Psi_0 (\vect{x},\vect{x}_{2-N}) 
\end{equation}
  is known as the conditional probability amplitude \cite{1975_Hunter_a}.  Evidently the function $\Phi^0_{N-1}(\vect{x},\vect{x}_{2-N})$ is normalized for every value of $\vect{x}$. For a given electron at $\vect{x}$, $|\Phi^0_{N-1}(\vect{x},\vect{x}_{2-N})|^2$  gives probability of  finiding other electrons at $\vect{x}_{2-N}$.
\par  For the corresponding Kohn-Sham system given by Eq. (\ref{I13}), the effective potential is known as the Pauli potential \cite{MARCH_1985,LEVY_1988} and is easily derived to  be \cite{LEVY_1988,Gritsenko_1994} ( see Appendix also )
 \begin{equation}
 \begin{split}
v_{eff}^{Pauli}[\{\phi_i ^0 \}](\vect{r})&= \frac{\sum_i(\epsilon_{max}-\epsilon_{i}) |\phi_i^0(\vect{r})|^2}{\rho_0(\vect{r})}  + \frac{\sum_i |\nabla \phi_i^0(\vect{r})|^2}{2 \rho_0(\vect{r})} \\
&-\frac{1}{8} \Big | \frac{\nabla \rho_0(\vect{r})}{\rho_0(\vect{r})}\Big|^2 
\end{split}
 \end{equation}
where $\epsilon_i$ are the eigenenergies of occupied orbitals, $\epsilon_{max}$ is the highest occupied orbital  eigenenergy and $\rho_0(\vect{r}) = \sum_i |\phi_i^0(\vect{r})|^2$ is the density. Note that this potential for single orbital systems is zero. In passing we note that this expression along with $v_{eff}(\vect{r})$ for Hartree-Fock wavefunction (see Appendix )
has been used \cite{Nagy_1997} in the past to derive the KLI approximation \cite{KLI_1992} to the exchange-only optimized potential \cite{Talman_1978}. The exchange-correlation potential $v_{xc}[\rho_0](\vect{r})$ appearing in the Kohn-Sahm equation  is given in terms of these effective potentials as 
\begin{equation}
\begin{split}
v_{xc}[\rho_0](\vect{r})& = v_{xc}^{\Psi_0}(\vect{r})  \\ &=  v_{eff}[\Psi_0](\vect{r})  - v_{eff}^{Pauli}[\{\phi_i ^0 \}](\vect{r})  -v_H(\vect{r})\label{vxc_psi}.
\end{split}
\end{equation}
Note that for a given ground-state wavefunction $\Psi_0$, the Kohn-Sham system is not known a priori so the exchange-correlation potential is obtained by solving the Kohn-Sham equation iteratively starting from an approximate $v_{xc}^{\Psi}(\vect{r})$ or equivalently $v_{eff}^{Pauli}(\vect{r})$. The Pauli potential and therefore the exchange-correlation potential improve with each iterative step. 
\par The presentation above has been in terms of the ground-state wavefunction $\Psi_0(\vect{x}_{1-N})$ and  the associated ground-state density $\rho_0(\vect{r})$. The question is what result will one get if an approximate wavefunction $\Psi(\vect{x}_{1-N})$ is employed in place of $\Psi_0(\vect{x}_{1-N})$ in the scheme presented above to calculate the exchange-correlation potential  $v_{xc}^{\Psi}(\vect{r})$  for the same external potential. This is the  approach  taken by Staroverov et al. (see section \ref{lpS_staroverov_cmpr}) who have calculated the exchange-correlation potential taking $\Psi$ to be the Hartree-Fock wavefunction (expressed in terms of Gaussian orbitals) or correlated wavefunctions calculated again using Gaussian basis-set. As noted earlier, they find that the exchange-correlation potential $v_{xc}^{\Psi}$  calculated is very close to the true exchange-correlation potential $v_{xc}^{\Psi_0} =v_{xc}[\rho_0] = v_{xc}[v_{ext}] $ in contrast to the exchange-correlation potential $v_{xc}^{\rho}$ calculated by inverting the corresponding density. As commented above  $v_{xc}^{\rho}(\vect{r})$ contains large oscillations near the nucleus and grows exponentially in the asymptotic regions. In the following we show that result obtained by Staroverov et al. are of general nature.  Thus if an approximate wavefunction $\Psi$ corresponding to the Hamiltonian of Eq. (\ref{I11b}) -for example that obtained by applying the  variational - is employed  in Eq. (\ref{LPS_mb}), the effective  potential $ v_{eff}[\Psi](\vect{r})$ so obtained is close to the true effective potential $v_{eff}[\Psi_0](\vect{r})$. Consequently the density $ \rho(\vect{r})$  calculated  by Eq. (\ref{scheq_den})  should also be close to the true density and prescription above  should lead to the exchange-correlation potential which  approximates the potential $v_{xc}[\rho_0] = v_{xc}[v_{ext}] $ well. We show this in the following with example of two-electron atom and ions.
\begin{figure*}
\begin{center}		  			
\includegraphics[scale=0.80]{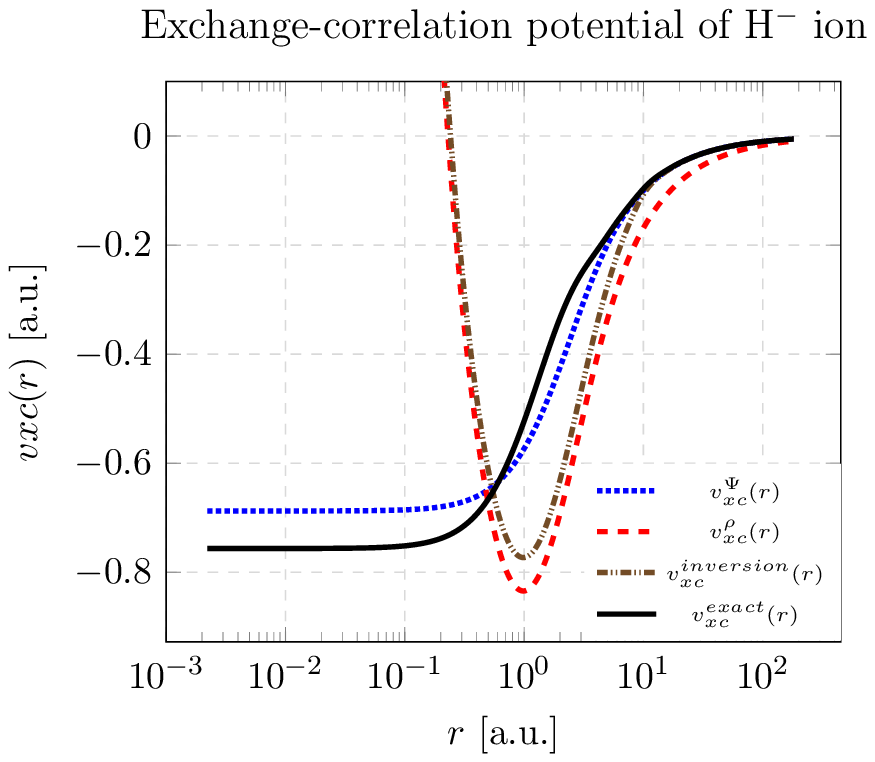} \hfil
\includegraphics[scale=0.80]{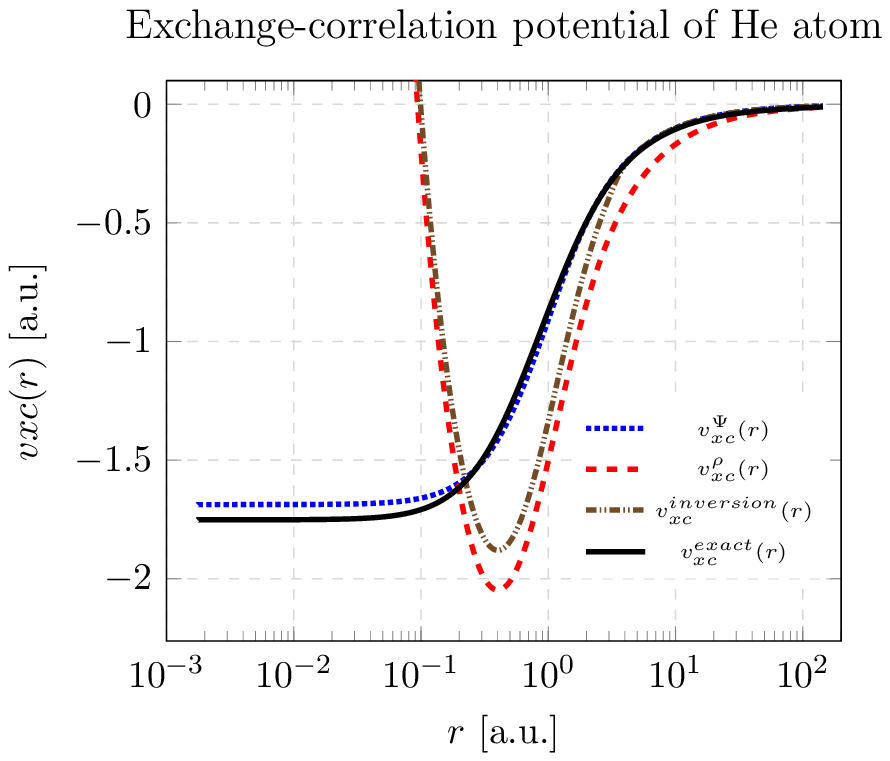}
\caption{\label{Fig1} Exchange-correlation potential for $\text{H}^{-}$ ion and He atom  using the product wavefunction given in Eq. (\ref{pwvfn}). It is evident that the exchange-correlation $v_{xc}^{\Psi}(\vect{r})$ obtained using wavefunction  is quite close to the exact potential $v_{xc}^{exact}(\vect{r})$ while the potential $v_{xc}^{\rho}(\vect{r})$ calculated using density inversion  deviates significantly near the nucleus. }
\end{center}
\begin{center}
 \includegraphics[scale=0.80] {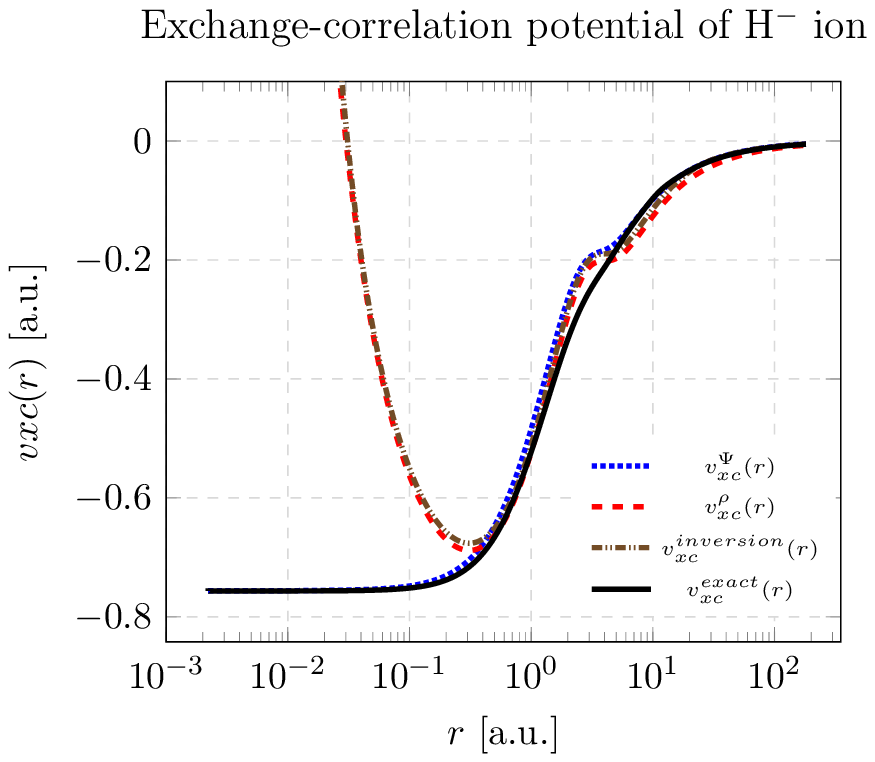} \hfil
\includegraphics[scale=0.80]{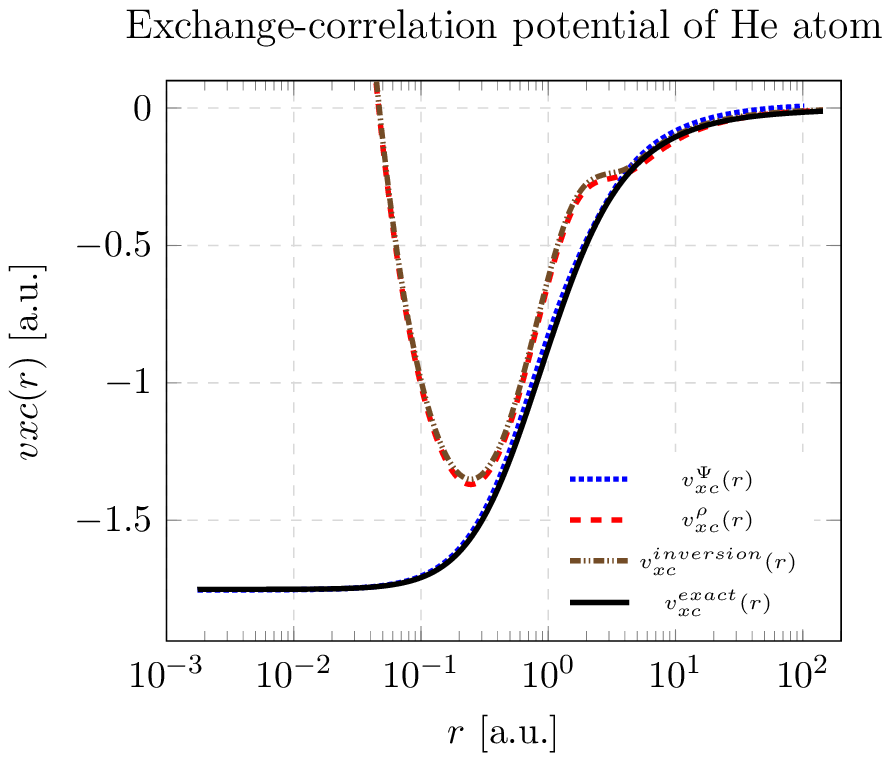}
\caption{\label{Fig2} Exchange-correlation potential for $\text{H}^{-}$ ion and He atom  for the wavefunction given in Eq. (\ref{cswvfn}). Again the exchange-correlation $v_{xc}^{\Psi}(\vect{r})$  obtained using wavefunction   is quite close to  the exact potential $v_{xc}^{exact}(\vect{r})$ while the potential $v_{xc}^{\rho}(\vect{r})$ calculated using density inversion  diverges near the nucleus.}
\end{center}
\begin{center}
\includegraphics[scale=0.80]{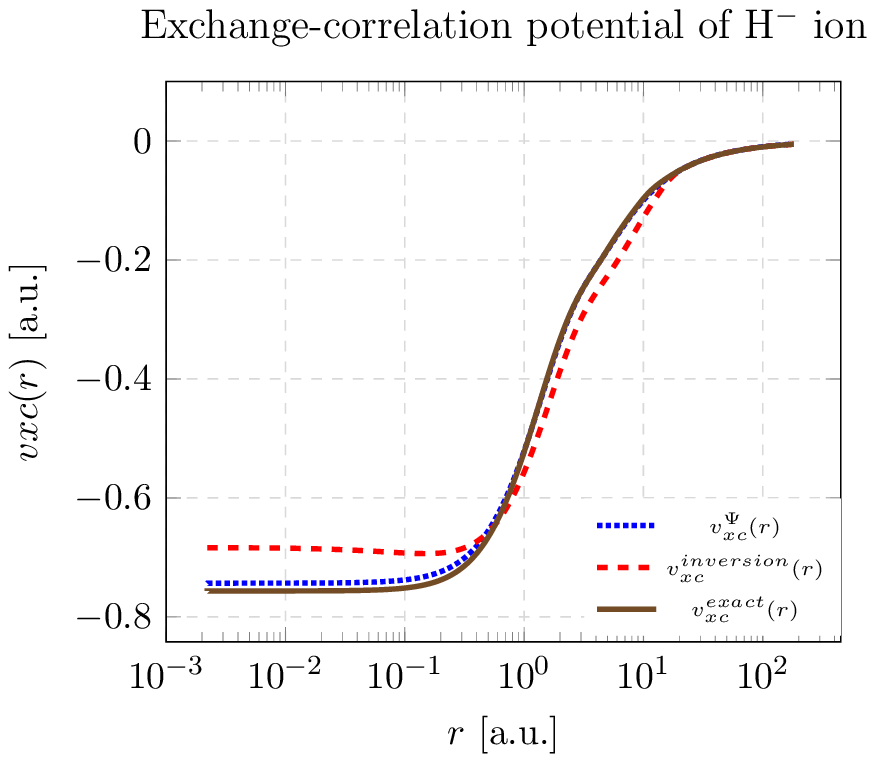}  \hfil
\includegraphics[scale=0.80]{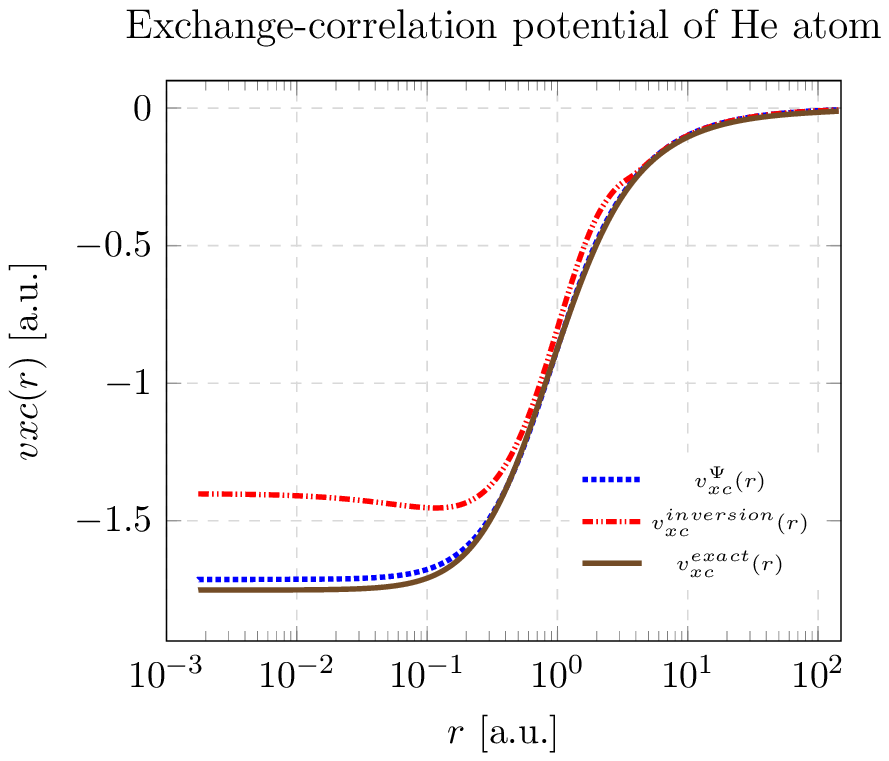}  
\caption{\label{Fig3} Exchange-correlation potential for $\text{H}^{-}$ ion  and He atom for  Le Sech-wavefunction. Again while the wavefunction gives energy close to exact result but exchange-correlation potential $v_{xc}^{inversion}(\vect{r})$ obtained using density inversion deviates significantly from exact one. However $v_{xc}^{\Psi}(\vect{r}) $ calculated using wavefunction is quite close to exact potential .}
\end{center}
\end{figure*} 
 \subsection{Results of applying LPS expression to obtain $v_{xc}(\vect{r})$ using approximate wavefunctions}\label{Results}
 In this section we describe the results of applying the  LPS equation to obtain the  exchange-correlation potential from variationally optimized approximate  wavefunctions for two-electron interacting systems. These results indicate that even with these wavefunctions, the LPS expression leads to accurate exchange-correlation potentials. On the other hand,  inversion of the corresponding densities gives potentials that are quite different from the exact ones. Our results are then connected to the work of Staroverov et al.\cite{Viktor_2013,Viktor_2015,Viktor_2015_JCP,Ospadov_2017}  who have obtained highly accurate exchange-correlation potential for atoms and molecules using wavefunction expressed in terms of finite Gaussian basis set. While methods based on the direct inversion of density in such cases give rise to wild oscillations in the exchange-correlation potential \cite{Mura_jcp_1997,savin_2003,2011_Jacob_jcp,2013_jacob_jcp,Staroverov_2017b}, the use of wavefunction yields highly accurate exchange-correlation potential. In the following we show through the examples of two-electron atoms and ions that the LPS expression leads to well behaved exchange-correlation  potentials for  approximate wavefunctions in  general.
\par We start with the example of optimized product wavefunction 
\begin{equation}
 \Psi(\vect{r},\vect{r}') = \phi_a(\vect{r})\phi_a(\vect{r}') \label{pwvfn}
\end{equation}  
for interacting Hamiltonian, where 
\begin{equation}
 \phi_a(\vect{r}) = \sqrt{\frac{a^3}{\pi}} e^{-ar} \label{var_orb}
 \end{equation}
   with $a= Z-\frac{5}{16} $, energy $ = -a^2$ and apply it to obtain the exchange-correlation potential. As shown below, it can be calculated analytically.
\par  The LPS effective potential corresponding to the product wavefuction given in Eq. (\ref{pwvfn}) is (up to a constant, constant is so chosen that potential goes to zero as $\vect{r}  \to \infty$)
\begin{equation}
 v_{eff}[\Psi](\vect{r}) = \frac{1}{2} \int\frac{\rho(\vect{r}')}{|\vect{r}'-\vect{r}|}d\vect{r}'  , 
\end{equation}
where $\rho(\vect{r}) = 2 |\phi_a(\vect{r})|^2 = \frac{2a^3}{\pi}e^{-2ar}$ is the electronic density of the system. Thus the exchange-correlation potential is given as
\begin{equation}
v_{xc}^{\Psi}(\vect{r}) = -\frac{1}{2} \int\frac{\rho(\vect{r}')}{|\vect{r}'-\vect{r}|}d\vect{r}'  = -\frac{1}{r} + \frac{e^{-2ar}}{r}(1+ar) . \label{vx_pwfn_lps}
\end{equation}
 Note the  expression above in terms of the density is the same  as in Hartree-Fock theory for two electron systems. On the other hand, direct inversion of the density using Kohn-Sham equation gives (up to a constant) 
 \begin{equation}
 \begin{split}
v_{xc}^{\rho}(\vect{r}) &= \frac{1}{2}\frac{\nabla^2\phi_a}{\phi_a} +\frac{Z}{r} - \int\frac{\rho(\vect{r}')}{|\vect{r}'-\vect{r}|}d\vect{r}' \\
& =  \frac{(Z-a-2)}{r} + \frac{2e^{-2ar}}{r}(1+ar).
 \label{vx_pwfn_density}
 \end{split}
 \end{equation}
 As is clearly seen from the expressions above, there is a significant difference between the two potentials. This is displayed in Fig. (\ref{Fig1}) where the potentials obtained in Eq. (\ref{vx_pwfn_lps}) and Eq. (\ref{vx_pwfn_density}) are plotted for the $\text{H}^{-}$ ion and He atom.  Also plotted in Fig. (\ref{Fig1}) is the exact exchange-correlation potential \cite{Umrigar_PRA_1994} for the $\text{H}^{-}$ ion and He atom. It is evident that exchange-correlation potential obtained using density inversion $v_{xc}^{\rho}(\vect{r})$ deviates significantly from the exact potential $v_{xc}^{exact}(\vect{r})$ and diverges near the nucleus. However, the potential $v_{xc}^{\Psi}(\vect{r})$ obtained using wavefunction   is close to $v_{xc}^{exact}(\vect{r})$ \cite{Umrigar_PRA_1994}. Equally important,  $v_{xc}^{\Psi}(\vect{r})$ has the same shape as the exact potential.  Fig. (\ref{Fig1})  also shows the exchange-correlation potential $v_{xc}^{inversion}(\vect{r})$ obtained numerically using density-to-potential inversion algorithm. For this we have used the hybrid method given  in our recent work \cite{Kumar_2019}.   We point out that in principle  $v_{xc}^{\rho}(\vect{r})$ and $v_{xc}^{inversion}(\vect{r})$ should be exactly the same but are slightly different from each other due to numerical implementation of the inversion algorithm. The potential $v_{xc}^{inversion}(\vect{r})$ is close to $v_{xc}^{\rho}(\vect{r})$ and shows divergent behavior near the nucleus.
 \par Having shown that the two results for the exchange-correlation potential are significantly different  for the product wavefunction, next we consider a correlated wavefunction that has the form \cite{bs_2014}(with optimization parameters $a$ and $b$)
\begin{equation}
\Psi(\vect{r},\vect{r}') = C_N \Big(e^{-ar}e^{-br'} +e^{-ar'}e^{-br}\Big). \label{cswvfn}
\end{equation}
 Here 
\begin{equation}
C_N =  \frac{1}{\pi} \Bigg[\frac{1}{ \big( \frac{2}{a^3b^3} +\frac{128}{(a+b)^6}\big)}\Bigg]^{1/2}
\end{equation}
is the normalization constant. The parameters $a$ and $b$ are optimized by minimizing the expression for the total energy
 \begin{equation}
E(a,b) = (E_K+E_{nucl}+E_{int}),
\end{equation}
where
\begin{equation}
E_K = \pi^2C^2_N\Big(\frac{a^2+b^2}{a^3b^3} +\frac{128ab}{(a+b)^6}\Big),\\
\end{equation}
\begin{equation}
E_{nucl} = -\pi^2C^2_N Z\Big(\frac{2(a+b)}{a^3b^3} +\frac{128}{(a+b)^5}\Big),\\
\end{equation}
\begin{equation}
 E_{int} = \pi^2C^2_N\Big(\frac{2(a^2+b^2 +3ab)}{a^2b^2(a+b)^3} +\frac{40}{(a+b)^5}\Big)
 \end{equation}

are  the kinetic, nuclear  and the  electron-electron interaction energies, respectively. For H$^{-}$ ion and He atom , the optimized values of parameters $(a,b)$ are $(1.0392, 0.2832 )$   and  $(2.1832, 1.1885)$,  respectively. The corresponding energies are $(-0.5133,-2.8756)$ Hartree, respectively.
\par This is again a wavefunction where expressions for various quantities and those for $v_{xc}^{\Psi}(\vect{r})$ and $v_{xc}^{\rho}(\vect{r})$ can be derived analytically. Those for different components of  the total energy have been given above. For the other relevant quantities - the density $\rho(\vect{r})$, Hartree potential $v_H(\vect{r})$, $ v_{eff}[\Psi](\vect{r})$, $v_{xc}^{\Psi}(\vect{r})$, $v_{xc}^{\rho}(\vect{r})$ -  the expressions are:
  \begin{equation}
\rho(\vect{r}) = 2\pi C_N^2\Big[ \frac{e^{-2ar}}{b^3} + \frac{e^{-2br}}{a^3}
+ \frac{16 e^{-(a+b)r}}{(a+b)^3}\Big]
 \end{equation}
\begin{equation}
\begin{split}
v_H (\vect{r})&= \frac{2\pi^2 C_N^2}{r}\Big[ \frac{2 -(1+ar)e^{-2ar} -(1+br)e^{-2br}}{a^3b^3} 
 \\
&+\frac{128 \{1 -(1+\frac{(a+b)r }{2})e^{-(a+b)r}\}}{(a+b)^6}  \Big] \label{vh_lps_psi}
\end{split}
\end{equation}
\begin{equation}
\begin{split}
v_{eff}&[\Psi](\vect{r})=  \frac{\pi C_N^2}{\rho(\vect{r})} \Bigg \{-\frac{2 e^{-2(a+b)r} }{r} \Big[ \frac{(1+ar)}{a^3} +\frac{(1+br)}{b^3}   \\
&+\frac{16(1+\frac{(a+b)r}{2})}{(a+b)^3}\Big] + \frac{1}{r}\Big[ \frac{e^{-2ar}}{b^3} +\frac{e^{-2br}}{a^3} 
+\frac{16e^{-(a+b)r}}{(a+b)^3}\Big]\\  
&+   \Big[ \frac{e^{-2ar}}{b} +\frac{e^{-2br}}{a} 
+ \frac{16 ab e^{-(a+b)r}}{(a+b)^3} \Big] \\
&-  2Z\Big[ \frac{e^{-2ar}}{b^2} +\ \frac{e^{-2br}}{a^2} 
+ \frac{8 e^{-(a+b)r}}{(a+b)^2} \Big] \\
&+   \Big[ \frac{a^2e^{-2ar}}{b^3} +\frac{b^2e^{-2br}}{a^3} 
+ \frac{16 ab e^{-(a+b)r}}{(a+b)^3} \Big] \Bigg\}\\
 &  -\frac{1}{8}\Big|\frac{\nabla \rho (\vect{r})}{ \rho(\vect{r})}\Big|^2 +\frac{Z^2}{2} \label{v_lps_psi}
\end{split}
\end{equation}
\begin{align}
 v_{xc}^{\Psi}(\vect{r})= v_{eff}[\Psi](\vect{r}) -v_H(\vect{r}) \label{vxc_lps_psi}.
\end{align}
On the other hand, the expression for  the exchange-correlation potential obtained from direct inversion of the density is
\begin{equation}
\begin{split}
v_{xc}^{\rho}(\vect{r})&= \frac{2\pi C_N^2}{\rho(\vect{r})}\Big[
  \frac{a^2 e^{-2ar}}{b^3 }+ \frac{b^2 e^{-2br}}{a^3}   
 + \frac{4 e^{-(a+b)r}}{(a+b)} \Big]  \\
 &+ \frac{2\pi C_N^2}{\rho(\vect{r})r}\Big[\frac{a e^{-2ar}}{b^3 }+
 \frac{b e^{-2br}}{a^3} + \frac{8 e^{-(a+b)r}}{(a+b)^2} \Big]   \\
 &- \frac{2\pi^2 C_N^4}{\rho^2(\vect{r})}\Big[\frac{a e^{-2ar}}{b^3 }
+ \frac{b e^{-br}}{a^3}    + \frac{8 e^{-(a+b)r}}{(a+b)^2}  \Big]^2   \\
 &-v_H(\vect{r}) +\frac{Z}{r} .      
 \end{split}
 \end{equation}
The potentials $v_{xc}^{\Psi}(\vect{r})$ and $v_{xc}^{\rho}(\vect{r})$  for $\text{H}^{-}$ ion and He atom are plotted in Fig. (\ref{Fig2}) along with the exact potential $v_{xc}^{exact}(\vect{r})$ calculated in ref. \cite{Umrigar_PRA_1994} and potential $v_{xc}^{inversion}(\vect{r})$ obtained numerically using inversion algorithm.  Again it is evident that $v_{xc}^{\Psi}(\vect{r})$ is close to and has the same shape as the exact potential $v_{xc}^{exact}(\vect{r})$. On the other hand 
 $v_{xc}^{\rho}(\vect{r})$ and $v_{xc}^{inversion}(\vect{r})$ both deviate substantially from $v_{xc}^{exact}(\vect{r})$ . As these wavefunctions are improved further, the exchange-correlation potential $v_{xc}^{\Psi}(\vect{r})$ becomes closer to $v_{xc}^{exact}(\vect{r})$. The potential $v_{xc}^{\rho}(\vect{r})$ also improves but may still remain different from the exact potential. For example for the Le Sech wavefunction \cite{Sech_JPB.30.L47,Chauhan_CPL.639.248,rabeet_1602.07042}, although energy is quite accurate but $v_{xc}^{inversion}(\vect{r})$  still remains different from $v_{xc}^{\Psi}(\vect{r})$ or  $v_{xc}^{exact}(\vect{r})$ (see Fig. (\ref{Fig3})).  In passing we note that the expressions given by Eqs. (\ref{vh_lps_psi}, \ref{v_lps_psi},\ref{vxc_lps_psi}) with the optimized the  $a$ and $b$ can be considered to be  reasonably good analytical expressions for the Hartree potential, LPS effective potential and the exchange-correlation potential for the He atom and isoelectronic positive ions. 
\par Besides the examples given above, we now describe results available in the literature. These consider approximate wavefunction expressed in finite basis set and construct \cite{Ospadov_2018,Staroverov_2017b,Viktor_2015,Viktor_2015_JCP,Viktor_2013} the exchange-correlation potential according to the details given in the section \ref{lpS_staroverov_cmpr} below. The potential so obtained is again found to be close to the true potential. 
\par As is clear from the discussion  above,  use of the LPS expression  leads to exchange-correlation potentials  which are close to the exact results. This is in contrast to those constructed by inversion of the density. In an extreme example, use of Gaussian basis in calculations give \cite{Schipper1997,2011_Jacob_jcp, 2013_jacob_jcp,Staroverov_2017b} wild oscillation in the potentials ; these can make the resulting potential deviate from the actual potential so much that there is no resemblance between the two. These oscillations have been attributed \cite{Schipper1997,savin_2003} to the Gaussian basis orbitals being the solution for  a harmonic oscillator potential. The question arises why the LPS expression leads to such accurate results. We answer this question in this paper by analyzing $  v_{eff}[\Psi](\vect{r})$ for approximate wavefunctions $\Psi $.  The approximate nature of the wavefunction may be due to the form chosen for it or due to the  use of finite basis set. The analysis is based on a comparison between $v_{eff}[\Psi](\vect{r})$ and  $v_{eff}[\rho](\vect{r})$, where the latter is obtained from the use of density directly in the LPS equation. The expression of $v_{eff}[\rho](\vect{r})$ is given below in Eq .(\ref{lps_den})
\section{Theory: Well behaved nature of $v_{eff}[\Psi](\vect{r})$ and $v_{xc}^{\Psi}(\vect{r})$ for approximate wavefunctions} 
The understanding of why the inversion of an approximate density generally leads to the exchange-correlation potential with large deviations from the exact one and why the LPS effective potential gives the exchange-correlation potential close to exact one can be summarized in one sentence: \textit { the external potential $\overline{v}_{ext}(\vect{r})$ corresponding to an approximate ground state density is different from the true external potential $v_{ext}(\vect{r})$ and this difference between the two potentials appears in the exchange-correlation potential.} Such a correlation between density and  potential  has been suggested earlier in a qualitative manner \cite{Schipper1997,savin_2003}. This is further supported by the observation \cite{2013_jctc_viktor} that the oscillations in the Kohn-Sham exchange-correlation potential obtained from the inversion of a density or equivalently the Kohn-Sham orbitals depend primarily on the  basis set used for the  calculation and is independent of the functional used for generating the density. In this section we prove the statement above mathematically. Furthermore, observing that the expression for the effective potential $ v_{eff}[\Psi](\vect{r})$ has the true external potential in it (see Eq. (\ref{lps wvfn}) below), we show that the difference between  $v_{eff}[\rho](\vect{r})$ and $ v_{eff}[\Psi](\vect{r})$  arises from the difference $ \nabla v_{ext}(\vect{r})$ between $\overline{v}_{ext}(\vect{r})$  and  $v_{ext}(\vect{r})$. 
 \par The LPS effective potential is given in terms of the density $\rho(\vect{r})$ as 
\begin{equation}
\begin{split}
v_{eff}[\rho] (\vect{r})&= \frac{1}{2}\frac{\nabla^2 \rho^{1/2}(\vect{r})}{\rho^{1/2}(\vect{r})} -v_{ext}(\vect{r}) +\mu \\
&= \frac{\nabla^2 \rho(\vect{r})}{4 \rho(\vect{r})} -\frac{1}{8}\Big|\frac{\nabla \rho(\vect{r})}{\rho(\vect{r})}\Big|^2  \\
 &~~~~~~~ -v_{ext}(\vect{r}) +E_N^0-E_{N-1}^0. \label{lps_den}
\end{split}
\end{equation}
Here we have used the fact \cite{PPLB} that $\mu = -$ionization potential $=E_N^0-E_{N-1}^0$, where $E_N^0$ and $E_{N-1}^0$ are the ground state energies of the $N$ and $N-1$ electron systems. It is clear from the equation above that the ratio of the gradient of density to the density and the ratio of the Laplacian of  density to the density determine the structure of $ v_{eff}[\rho](\vect{r})$ and that may lead to large deviations from the exact structure if  the density is approximate. For example, let us see what will happen if the density fails to satisfy the nuclear cusp condition \cite{Kato_CPAM.10.151} in an atom exactly i.e. $\frac{d\rho}{dr} \ne -2Z\rho$. In that case $\frac{1}{2}\frac{\nabla^2 \rho^{1/2}}{\rho^{1/2}}$ does not have the term $- \frac{Z}{r}$ to cancel $-v_{ext}(\vect{r})$ and therefore the effective potential diverges as $\frac{Z}{r}$ for $r \to 0$. This is what is seen in Fig. (\ref{Fig1}) and Fig. (\ref{Fig2}) for such wavefunctions. Consider another example where an orbital is expanded in terms of Gaussian orbitals. In that case for $r \to \infty$, only one Gausssian orbital will contribute to the density and $ v_{eff}[\rho] \propto r^2$ in that limit. Thus it is seen that the deviation from the exact LPS effective potential arises from the difference $\Delta v_{ext} (\vect{r}) = \overline{v}_{ext}(\vect{r}) - v_{ext}(\vect{r})$ in the external potential $\overline{v}_{ext}(\vect{r})$ corresponding to the approximate density (and the wavefunction) and the exact external potential $v_{ext}(\vect{r})$. We show this explicitly in the following. Notice that the maximum deviation occurs when the density is dominated by one orbital or one basis function.
 \par Consider the conditional probability amplitude \cite{1975_Hunter_a}
\begin{equation}
\Phi_{N-1}(\vect{x},\vect{x}_{2-N})= \Bigg(\frac{N}{\rho(\vect{r})}\Bigg)^{1/2} \Psi(\vect{x},\vect{x}_{2-N}) \label{phi_n-1_1}
 \end{equation}
 constructed from  the wavefunction $\Psi(\vect{x},\vect{x}_{2-N})$ and the  corresponding density $\rho(\vect{r})$.  If $\Psi$ is exact then the external potential is $v_{ext}(\vect{r})$ and if $\Psi$ is approximate the external potential is denoted as $\overline{v}_{ext}(\vect{r})$. We now proceed as follows. First one can easily show that 
  \begin{equation}
  \begin{split}
  \frac{1}{2}\frac{\nabla^2 \rho^{1/2}(\vect{r})}{\rho^{1/2}(\vect{r})} &= \frac{N}{2\rho(\vect{r})}\int \Psi^* \nabla^2\Psi d \sigma d\vect{x}_{2-N}  \\
  & + \frac{1}{2}\int |\nabla \Phi_{N-1} |^2d\sigma d\vect{x}_{2-N}. \label{phi_n-1_2}
  \end{split}
 \end{equation}
To derive this relation, start by calculating $\frac{1}{2}\int |\nabla \Phi_{N-1} |^2d\sigma d \vect{x}_{2-N}$  and rearrange  terms in the resulting expression. We now consider an approximate wavefunction $\Psi$. Then in the equation above the first term 
\begin{equation}
\begin{split}
\frac{N}{2\rho(\vect{r})} & \int \Psi^* \nabla^2\Psi d\sigma d\vect{x}_{2-N} \\
&= -\overline{E}^0_N  + \langle \Phi_{N-1}|\overline{H}_{N-1} |\Phi_{N-1}\rangle \\
&\quad +\overline{v}_{ext}(\vect{r}) 
+\int  \frac{\rho_{N-1}(\vect{r};\vect{r}')}{|\vect{r}-\vect{r}'|} d\vect{r}'
\end{split}
\end{equation}
by using the Schr\"{o}dinger equation $\overline{H}_N\Psi=\overline{E}_N\Psi$. Here 
\begin{equation}
\overline{H}_{N}(\vect{r}_{1-N}) = \sum_{i=1}^{N} \Big( -\frac{1}{2}\nabla^2_i +\overline{v}_{ext}(\vect{r}_i) \Big) +\frac{1}{2} \sum \limits_{ \underset{i \ne j}{i,j =1} }^{N} \frac{1}{|\vect{r}_i- \vect{r}_j|},
\end{equation}
\begin{equation}
\overline{H}_{N-1}(\vect{r}_{2-N}) = \sum_{i=2}^{N} \Big( -\frac{1}{2}\nabla^2_i +\overline{v}_{ext}(\vect{r}_i) \Big) +\frac{1}{2} \sum \limits_{ \underset{i \ne j}{i,j =2} }^{N} \frac{1}{|\vect{r}_i- \vect{r}_j|} \label{h_n_1}
\end{equation}
and  $\rho_{N-1}(\vect{r};\vect{r}')$ is the density of $N-1$ particle at $\vect{r}'$ calculated from $\Phi_{N-1}(\vect{x},\vect{x}_{2-N})$. Thus the LPS effective potential using Eq. (\ref{lps_den}) is given as 
\begin{equation}
\begin{split}
v_{eff}[\rho](\vect{r}) &= \int \frac{\rho_{N-1}(\vect{r};\vect{r}')}{|\vect{r}-\vect{r}'|} d\vect{r}' \\ & + \langle \Phi_{N-1}|\overline{H}_{N-1} -E^0_{N-1}|\Phi_{N-1}\rangle 
 \\ &+ \frac{1}{2}\int |\nabla \Phi_{N-1} |^2d \sigma d\vect{x}_{2-N} 
 + \Delta v_{ext} (\vect{r})  \\ &+ (E^0_N-\overline{E}^0_N), \label{vlps_phi_n-1}
\end{split}
\end{equation}
where we recall that  $\Delta v_{ext} (\vect{r}) = \overline{v}_{ext}(\vect{r}) - v_{ext}(\vect{r})$.
Notice that the expression for $v_{eff}[\rho](\vect{r})$  contains $\overline{v}_{ext}(\vect{r})$. On the other hand, $v_{eff}[\Psi](\vect{r})$ corresponding to $\Psi$ is evaluated as 
\begin{equation}
\begin{split}
v_{eff}[\Psi](\vect{r})&= \int \frac{\rho_{N-1}(\vect{r};\vect{r}')}{|\vect{r}-\vect{r}'|} d\vect{r}' \\
 & +\langle \Phi_{N-1}|H_{N-1} -E^0_{N-1}|\Phi_{N-1}\rangle  \\ 
& + \frac{1}{2}\int |\nabla \Phi_{N-1} |^2d \sigma d\vect{x}_{2-N},
 \label{lps wvfn}
\end{split}
\end{equation}
where $H_{N-1}$ is given by Eq. (\ref{h_n_1}) by replacing $\bar{v}_{ext}(\vect{r})$ by $v_{ext}(\vect{r})$. Thus the difference
 \begin{equation}
 \begin{split}
v_{eff}[\rho](\vect{r})- v_{eff}[\Psi](\vect{r})&  = (E^0_N-\overline{E}^0_N) + \Delta v_{ext} (\vect{r})  \\
&+ \int \Delta v_{ext} (\vect{r})\rho_{N-1}(\vect{r};\vect{r}')d\vect{r}' . \label{delta_lps_pot}
\end{split}
\end{equation}
As is clear from the expression above, the difference between the effective potentials calculated by inverting the density and that obtained from the  wavefunction using the LPS  expression arises from the difference between the external potential $\overline{v}_{ext}(\vect{r})$ corresponding to the approximate wavefunction $\Psi$ and the true external potential $v_{ext}(\vect{r})$. It is this difference that appears in the exchange-correlation potential  $v_{xc}^{\rho}(\vect{r})$ calculated from the density-to-potential inversion and $v_{xc}^{\Psi}(\vect{r})$ obtained using the effective potential $v_{eff}[\Psi](\vect{r})$ . 
\par  Having obtained the difference between the effective LPS potential $v_{eff}[\rho](\vect{r})$ calculated by inverting the density  and $v_{eff}[\Psi](\vect{r})$ by the use of wavefunction dependent expression, we now pay attention to the behavior of $v_{eff}[\Psi](\vect{r})$.  We focus on understanding whether it deviates from the exact potential by a large amount. For this expand the conditional probability amplitude as 
 \begin{equation}
 \Phi_{N-1}(\vect{x},\vect{x}_{2-N}) = \sum_i f_i(\vect{x}) \Psi^{i(0)}_{N-1}(\vect{x}_{2-N}) \label{fs}
 \end{equation}
where because of $\phi_{N-1}$ being normalized, $   \sum_i |f_i(\vect{x})|^2 =1$ so that $ |f_i(\vect{x})| \le 1$ for every value of $\vect{x}$; here $\Psi^{i(0)}_{N-1}$ are the eigenfunction for $(N-1)$ electron in the Hamiltonian with external potential $v_{ext}(\vect{r})$. We do this expansion so that $v_{eff}(\vect{r})$ is shown to be well behaved  independent of the expressions  given in Eq. (\ref{phi_n-1_1}) and Eq. (\ref{phi_n-1_2}).
\par For well behaved $\Psi$ \cite{schiff1955quantum,szabo1996modern,levine2009quantum} and because $\rho(\vect{r})$ for the ground-state is nonzero except when $r \to \infty$,  $ f_i$ and its gradient will be finite and smooth. Thus, all the terms in $v_{eff}[\Psi](\vect{r})$ viz. 
\begin{equation}
\begin{split}
 \int \frac{\rho_{N-1}(\vect{r};\vect{r}')}{|\vect{r}-\vect{r}'|} d\vect{r}' 
  &= (N-1)\sum_{i,j}\int f^*_i(\vect{x})f_j(\vect{x}) d\sigma  \\
   &\quad \int \frac{\Psi^{i(0)*}_{N-1}\Psi^{j(0)}_{N-1} }{|\vect{r}-\vect{r}'|}  d\vect{x}_{2-N},
\end{split}
\end{equation}
\begin{equation}
\begin{split}
\langle \Phi_{N-1}|H_{N-1} &-E^0_{N-1}|\Phi_{N-1}\rangle  \\
 &= \sum_{i \neq 0} (E^{i}_{N-1} -E^0_{N-1})  \int |f_i(\vect{x})|^2 d\sigma
\end{split}
\end{equation}
 and
 \begin{equation}
 \frac{1}{2}\int |\nabla \Phi_{N-1} |^2d \sigma d\vect{x}_{2-N} = \frac{1}{2} \sum_{i} \int |\nabla f_i(\vect{x})|^2 d\sigma
 \end{equation} 
 \begin{figure*}
 	\begin{center}
        \includegraphics[scale=0.80]{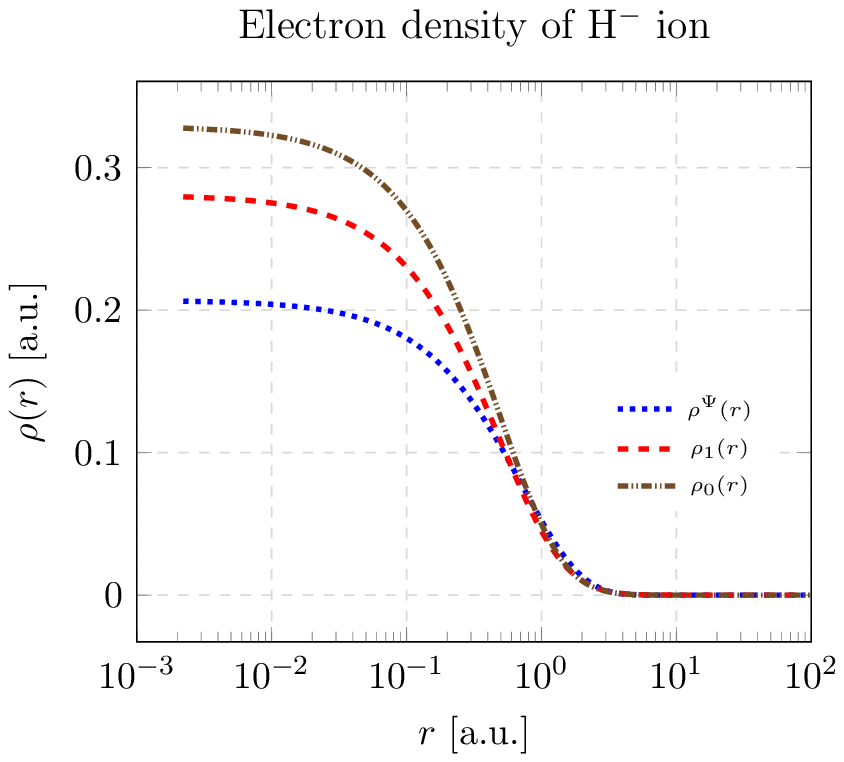} \hfil
 		\includegraphics[scale=0.80]{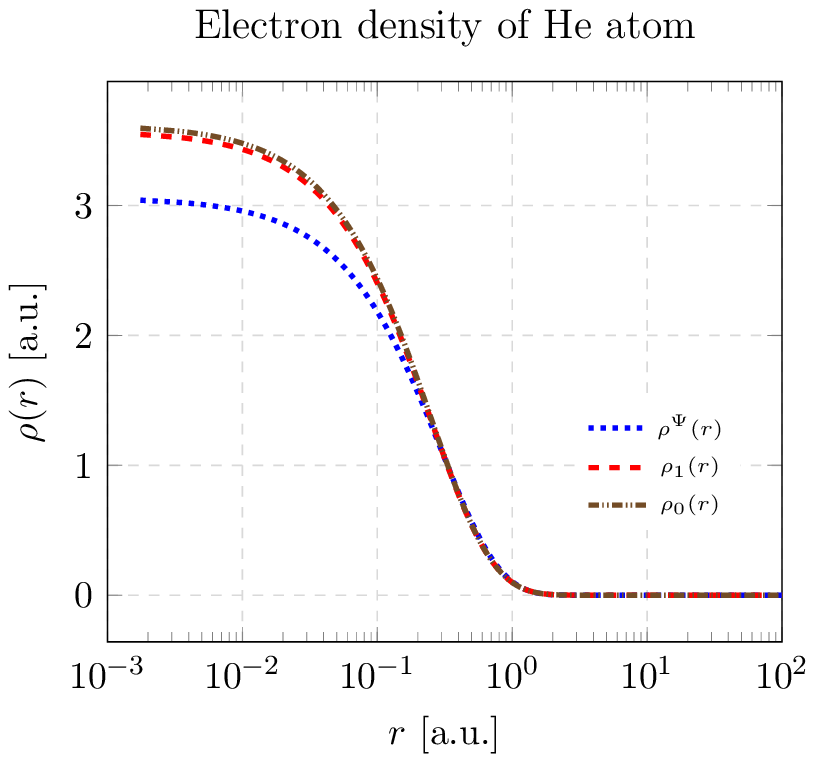}		  			
 		\caption{\label{Fig4} Electron density for  H$^{-}$ ion and He atom  using product wavefunction given in Eq. (\ref{pwvfn}). It can be seen that there is significant difference between  the density $\rho^{\Psi}$ obtained from wavefunction and exact density  $ \rho_0(\vect{r})$. While in comparison to  $\rho^{\Psi}$, the density $ \rho_{1}(\vect{r})$ obtained using LPS equation is more closer to exact density $ \rho_0(\vect{r})$ .
 		}
 	\end{center}
 \end{figure*} 
 \begin{figure*}
 	\begin{center}
        \includegraphics[scale=0.80]{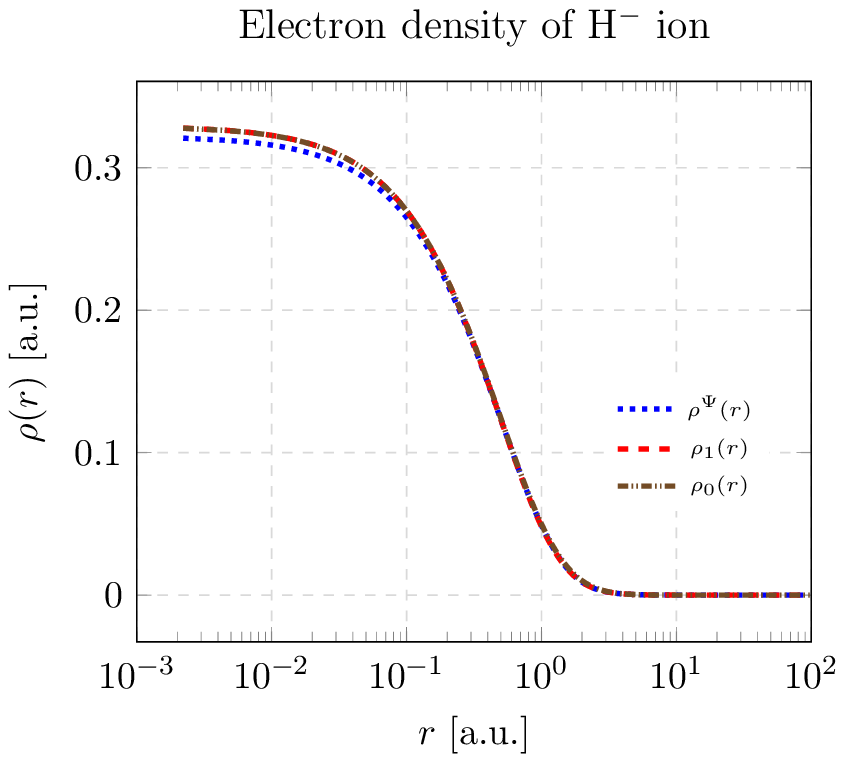} \hfil
 		\includegraphics[scale=0.80]{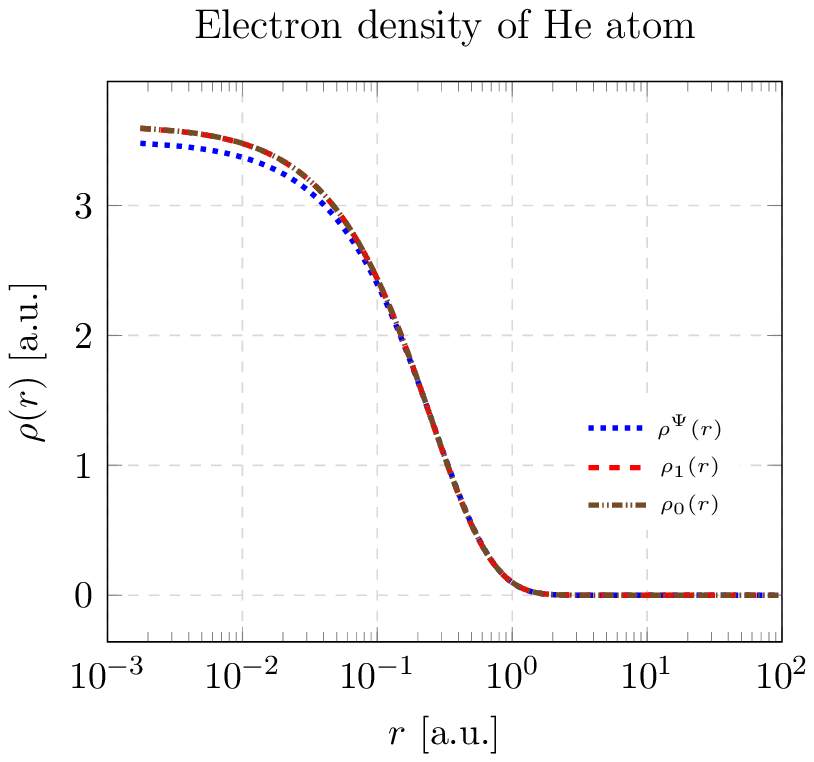}
 		\caption{\label{Fig5} Electron density for $\text{H}^{-}$ ion and He atom  for wavefunction given in Eq. (\ref{cswvfn}). Again the density $\rho_1(\vect{r})$  obtained by solving LPS equation having LPS potential calculated from wavefunction , is indistinguishably close to exact density $\rho _0(\vect{r})$.}
 	\end{center}
 \end{figure*} 
are also finite and do not become spuriously large . Notice that if in the expression above there  was a term having division by $|f|^2$ or $\rho(\vect{r})$, that term could have become large. Using the expression derived above, we now calculate deviation of $v_{eff}[\Psi] (\vect{r})$ from the exact one.
\par If $ f^0_i(\vect{x})$ are the functions corresponding to the exact wavefunction $\Psi_0$, then for  $f_i(\vect{x}) = f^0_i(\vect{x}) +\delta f_i(\vect{x})$ the effective potential can be written as
\begin{equation}
v_{eff}[\Psi] (\vect{r})= v_{eff}[\Psi_0](\vect{r})+\delta v_{LPS}(\vect{r})
\end{equation}
where 
\\
\begin{widetext}    
\begin{equation}
\begin{split}
\delta v_{LPS}&(\vect{r}) = \sum_{i \neq 0} (E^{i}_{N-1} -E^0_{N-1}) \int \Big(f^{0*}_i (\vect{x})\delta f_i(\vect{x}) +f^0_i (\vect{x})\delta f^*_i(\vect{x}) +|\delta f_i(\vect{x})|^2\Big) d\sigma \\
&  + (N-1) \sum_{i,j}\int \Big(\delta f^{*}_i(\vect{x}) f^0_j(\vect{x})+ f^{0*}_i(\vect{x})\delta f_j(\vect{x}) +
\delta f^*_i(\vect{x})\delta f_j(\vect{x})  \Big) d\sigma  
\int \frac{\Psi^{i(0)*}_{N-1}\Psi^{j(0)}_{N-1} }{|\vect{r}-\vect{r}'|}  d\vect{x}_{2-N}  \\
 & +\frac{1}{2}\sum_{i} \int \Big(\nabla  f^{0*}_i(\vect{x}) \cdot \nabla \delta f_i(\vect{x}) +\nabla  f^{0}_i(\vect{x}) \cdot \nabla \delta f^*_i(\vect{x}) +|\nabla  \delta f_i(\vect{x})|^2  \Big) d\sigma .\label{del_vlps}
 \end{split}
 \end{equation} 
 \end{widetext}
As is apparent   $\delta v_{LPS}(\vect{r})$    has terms that do not grow large erroneously. Furthermore, for small $\{ \delta f_i \}$, the difference is linear in $\{ \delta f_i \}$ . Thus it can be safely concluded that $ v_{eff}[\Psi](\vect{r})$   is close to   $  v_{eff}[\Psi_0](\vect{r})$  . The next question that arises is  about the behavior of the corresponding density and the exchange-correlation potential. We now address that.
\par The density corresponding to $v_{eff}[\Psi](\vect{r})$ is obtained by solving the LPS equation \cite{LPS_1984}
\begin{equation}
 \Big [-\frac{1}{2} \nabla^2+v_{ext}(\vect{r})+v_{eff}[\Psi](\vect{r}) \Big ]\rho^{1/2}(\vect{r}) = \mu\rho^{1/2}(\vect{r}) .
\end{equation}
Let us call this density $\rho_1(\vect{r})$. Since $v_{eff}[\Psi](\vect{r}) \approx v_{eff}[\Psi_0](\vect{r})$,  density $\rho_1(\vect{r})$ will be closer to the exact density $\rho_0(\vect{r})$ (given by $\Psi_0$) than the density $\rho^{\Psi}(\vect{r})$ corresponding to the approximate wavefunction $\Psi$ used. This is shown in Fig. (\ref{Fig4}) and Fig. (\ref{Fig5}) for H$^{-}$ ion and He atom using the product and  correlated wavefunctions given in Eq. (\ref{pwvfn}) and Eq. (\ref{cswvfn}), respectively. In these figures we have plotted $\rho_1(\vect{r})$ and $\rho^{\Psi}(\vect{r})$ associated with these wavefunctions. Also plotted is the exact density $\rho_0(\vect{r})$ \cite{Koga_IJQC.46.689}. We see that in all the cases, $\rho_1(\vect{r})$ is much closer to $\rho_0(\vect{r})$ in comparison to $\rho^{\Psi}(\vect{r})$(The maximum deviation is when product wavefunction is used for $\text{H}^{-}$ ion). This then also suggests a possible method of obtaining accurate ground state densities using an approximate wavefunction. This will be explored in the future. Note that for two electron systems Pauli potential $v^{Pauli}_{eff}$  is zero and therefore the LPS effective potential for two electron systems contains only the Hartree and the exchange-correlation potential.
\par Next, we observe the following. Since $v_{eff}[\Psi](\vect{r})$ is free from spuriously large deviations from  $v_{eff}[\Psi_0](\vect{r})$, it is anticipated that $v_{xc}^{\Psi}(\vect{r})$ will also not deviate much from $v_{xc}^{\Psi_0}(\vect{r})$. We now show this to be the case irrespective of whether the Kohn-Sham calculation is done exactly (fully numerically on a grid) or by employing a finite basis set, as long as a wavefunctional (orbital) based expression is used for the calculation of Pauli potential.
\par The exchange-correlation  potential $v_{xc}^{\Psi}(\vect{r})$ is obtained  from $v_{eff}[\Psi](\vect{r})$ by subtracting from it   $v^{Pauli}_{eff}[\{\phi_i \}](\vect{r})$ i.e. the Pauli potential for the Kohn-Sham orbitals . The convergence towards true $v_{xc}(\vect{r})$  is done by iterative process. The resulting potential $v_{xc}^{\Psi}(\vect{r})$  will be smooth and will not have erroneously large deviation from the exact potential if the Pauli potential  $v^{Pauli}_{eff}[\{\phi_i \}](\vect{r})$ is well behaved at each iteration for approximate $\{\phi_i \}$. That this is the case can be shown exactly in the same manner as done above for $v_{eff}[\Psi](\vect{r})$ . For this, at the $n^{th}$ iteration, we write
 \begin{equation}
 \Phi^{KS}_{N-1}(\vect{x},\vect{x}_{2-N}) = \sum_i f^{KS (n)}_i(\vect{x}) \Psi^{i,KS (n)}_{N-1}(\vect{x}_{2-N}),
 \end{equation}
where $\Psi^{i,KS (n)}_{N-1}(\vect{x}_{2-N})$ represent the determinant for $i^{th}$ exited state of the KS Hamiltonian corresponding to the $n^{th}$ iteration and $\{f^{KS (n)}_i\}$ have the same properties as in the fully interacting case (see Eq. (\ref{fs}) above). Then during each iteration
\begin{equation}
\begin{split}
 v_{eff}^{Pauli,(n)}&[\{\phi_i \}](\vect{r}) =  \\
  & \sum_{i \neq 0} (E^{i,KS(n)}_{N-1} -E^{0,KS(n)}_{N-1})\int |f^{KS(n)}_i(\vect{x})|^2 d\sigma  \\
& \quad \quad +\frac{1}{2} \sum_{i} \int |\nabla f^{KS(n)}_i(\vect{x})|^2 d\sigma  .
\end{split}
\end{equation}
Thus if we start iterations with a reasonable approximation (say LDA) to $v_{xc}$, this potential is always going to be free of erroneous large terms and close to the exact Pauli potential for the $n^{th}$ iteration,  thereby leading to a smooth exchange-correlation potential as iterations proceed. Furthermore, as shown in the Eq. (\ref{del_vlps}) , this exchange-correlation potential will be close to the true exchange-correlation potential after convergence.
\section{Analysis of  Ryabinkin, Kohut, and Staroverov (RKS) \cite {Viktor_2013, Viktor_2015} \& modified RKS (\texorpdfstring{\MakeLowercase{m}RKS}{mRKS}) \cite{Ospadov_2017} methods}
\label{lpS_staroverov_cmpr}
Having shown that the LPS potential calculated from wavefunction is well behaved and close to the true  potential, we now use this to develop an understanding of why the method of Ryabinkin, Kohut,  Staroverov (RKS) \cite {Viktor_2013, Viktor_2015} and modified RKS (mRKS)  \cite{Ospadov_2017} method give the accurate exchange-correlation potential and mRKS improves the RKS further.
\par The main equation used by Staroverov et al. is
\begin{equation}
\begin{split}
v_{xc}(\vect{r}) &= v^{\Psi}_s(\vect{r}) +\frac{\tau^{\Psi}(\vect{r})}{\rho(\vect{r})}- \frac{\nabla^2 \rho(\vect{r})}{4 \rho(\vect{r})}-\epsilon^{\Psi}(\vect{r}) \\
&  \quad \quad -\frac{\tau^{KS}(\vect{r})}{\rho^{KS}(\vect{r})}+  \frac{\nabla^2 \rho ^{KS}(\vect{r})}{4\rho^{KS}(\vect{r})} +\epsilon^{KS}(\vect{r}).
   \label{rks_main}
\end{split}
\end{equation}
The quantities  on the right side of above equation are given in terms of  many-body wavefunction $\Psi$  and corresponding density $\rho(\vect{r})$. Here 
\begin{equation}
v^{\Psi}_s(\vect{r}) = \int \frac{\rho_{xc}(\vect{r},\vect{r}')}{|\vect{r}-\vect{r}'|} d\vect{r}',
\end{equation}
 \begin{equation}
\tau^{\Psi}(\vect{r}) = \frac{1}{2}
\int |\nabla \Psi |^2 d \sigma d\vect{x}_{2-N},
\end{equation}
and 
\begin{equation}
\epsilon^{\Psi}(\vect{r})=  \frac{N}{\rho(\vect{r})}\int  \Psi^* (E_N-H_{N-1}) \Psi d \sigma d\vect{x}_{2-N}
\end{equation}
are  potential of Fermi-Coulomb hole $\rho_{xc}(\vect{r},\vect{r}')$ , kinetic energy density and average energy respectively. Similarly quantities are obtained from  Kohn-Sham orbitals  $\{\phi_i^{KS}\}$ with corresponding eigenenergies  $\{\epsilon_i^{KS}\}$ are the kinetic energy density $\tau^{KS}(\vect{r}) = \frac{1}{2}
\sum_{i} |\nabla \phi^{KS}_i (\vect{r})|^2 $ and  the average ionization energy
$\epsilon^{KS}(\vect{r}) =  \frac{1}{\rho^{KS}(\vect{r})}\sum_{i}\epsilon_i^{KS} | \phi^{KS}_i (\vect{r})|^2$.
\par In their first algorithm for construction of exchange-correlation potential,  Ryabinkin, Kohut, and Staroverov (RKS) \cite {Viktor_2013, Viktor_2015} took the term $ \nabla^2 \rho ^{KS}(\vect{r}) / \rho^{KS}(\vect{r}) -  \nabla^2 \rho(\vect{r}) / \rho(\vect{r})$ in  Eq. (\ref{rks_main}) to be zero. Then  the equation for updating of the exchange-correlation potential becomes 
\begin{equation}
v^{RKS}_{xc}(\vect{r}) =v^{\Psi}_s(\vect{r}) +\frac{\tau^{\Psi}(\vect{r})}{\rho(\vect{r})}-\epsilon^{\Psi}(\vect{r})
  -\frac{\tau^{KS}(\vect{r})}{\rho^{KS}(\vect{r})}+\epsilon^{KS}(\vect{r}) . \label{rks}
\end{equation}
To apply the above equation one starts with a trial $\{\phi^{KS}_i,\epsilon^{KS}_i \}$ and the corresponding $v^{RKS}_{xc}(\vect{r}) $ is used in the Kohn-Sham equation
\begin{equation}
\Big[-\frac{\nabla^2}{2} +v_{ext}(\vect{r}) +v_H[\rho](\vect{r})+v^{RKS}_{xc}(\vect{r}) \Big]\phi^{KS}_i(\vect{r}) = \epsilon_i ^{KS}\phi^{KS}_i(\vect{r})
\end{equation}
to get the next set of Kohn-Sham orbitals and eigenenergy. This procedure  is applied until convergence condition of Kohn-Sham density obtained during two consecutive iterations is achieved. In the basis-set limit calculation the above equation gives $\rho^{KS}(\vect{r}) =\rho(\vect{r})$  and the resulting potential is the true exchange-correlation potential of density $\rho(\vect{r})$. For a finite basis set calculation the $\rho^{KS}(\vect{r}) \neq \rho(\vect{r}) $ and the resulting potential is an approximation to the true potential  conjugate to density $\rho(\vect{r})$. The  exchange-correlation potential obtained from other existing popular  density-to-potential inversion methods  \cite{Werden, Stott_1988,Gorling_1992, Zhao_1992,Wang_1993, Zhao_1993, Zhao_1994, Wang_1993, Vlb_1994, Schipper1997, WY2,Peirs_2003, Stott_2004, Wagner_2014,Hollins_2017,Wasserman_2017,Finzel2018,Kumar_2019}  depends upon what type of density it corresponds to and for a basis-set density it could show unphysical behavior. However, exchange-correlation potential obtained  by RKS method is found to be free from such pathological features. According to Staroverov et al. \cite{Ospadov_2017} the RKS method gives good results because by  taking $ \nabla^2 \rho ^{KS}(\vect{r}) / \rho^{KS}(\vect{r}) -  \nabla^2 \rho(\vect{r}) / \rho(\vect{r})= 0$ one sets it to its basis set limit value even if $\rho^{KS}(\vect{r}) \neq \rho(\vect{r})$ so the resulting exchange-correlation potential get close to its basis set limit. We point out that apart from imposing basis-set limit value on few quantities, it is the use of wavefunction dependent quantities in RKS method  which  play important role in giving the proper structure to the exchange-correlation potential obtained from it. This becomes transparent by writing  Eq. (\ref{rks}) in terms of the LPS potential $v_{eff}(\vect{r})$. Now using relations
\begin{equation}
\int \frac{\rho_{N-1}(\vect{r};\vect{r}')}{|\vect{r}-\vect{r}'|} d\vect{r}' = v_s^\Psi(\vect{r}) +v_H[\rho](\vect{r}),
\end{equation}
\begin{equation}
 \langle \Phi_{N-1}|H_{N-1}-E^0_{N-1} |\Phi_{N-1}\rangle = \mu - \epsilon^{\Psi}(\vect{r}),
\end{equation}
and
\begin{equation}
\frac{1}{2}\int |\nabla \Phi_{N-1} |^2d\sigma d\vect{x}_{2-N}= \frac{1}{\rho(\vect{r})}\Bigg( \tau^{\Psi}(\vect{r})
-\frac{1}{8}\frac{|\nabla \rho(\vect{r})|^2}{\rho(\vect{r})}\Bigg)
\end{equation}
the Eq. (\ref{rks}) is written (up to a constant) as 
\begin{equation}
\begin{split}
v_{xc}^{RKS}(\vect{r}) &=  v_{eff}[\Psi](\vect{r}) + \frac{1}{8}\Big|\frac{\nabla \rho(\vect{r})}{\rho(\vect{r})}\Big|^2   \\
&-  v_{eff}[\{\phi_i^{KS} \}](\vect{r})- \frac{1}{8} \Big|\frac{\nabla \rho_{KS}(\vect{r})}{\rho_{KS}(\vect{r})}\Big|^2  -v_H[\rho](\vect{r})  \label{rks_lps}.
\end{split}
\end{equation}
From the equation above it is seen that the RKS method utilizes the LPS potential written in terms of wavefunction for construction of exchange-correlation potential. Since we have shown that the  LPS potential $v_{eff}(\vect{r})$ obtained from many-body wavefunction $\Psi$  and Kohn-Sham orbitals $\{ \phi^{KS}_i\}$ is well behaved, so the resulting exchange-correlation potential obtained from RKS method is also expected to show proper structure. However, Eq. (\ref{rks_lps})  also contains density dependent term $|\nabla \rho(\vect{r}) / \rho(\vect{r})|^2 -| \nabla \rho_{KS}(\vect{r}) / \rho_{KS}(\vect{r})|^2$ in it whose effect may appear in the resulting potential. For  $\rho^{KS}(\vect{r}) \approx \rho(\vect{r})$ the contribution of density dependent term is vanishingly small. However, for the finite basis set calculation $\rho^{KS}(\vect{r}) \neq \rho(\vect{r})$ and the quantity $|\nabla \rho(\vect{r}) / \rho(\vect{r})|^2 -| \nabla \rho_{KS}(\vect{r}) / \rho_{KS}(\vect{r})|^2$ may give significant contribution and the resulting potential could have pathological features. This is seen for Ar atom  \cite{Ospadov_2017} where  exchange-correlation potential shows well behaved nature only for a  large basis-set calculation. However, by taking the $|\nabla \rho(\vect{r}) / \rho(\vect{r})|^2 -| \nabla \rho_{KS}(\vect{r}) / \rho_{KS}(\vect{r})|^2 = 0$ the above equation becomes
\begin{equation}
v_{xc}^{m,RKS}(\vect{r}) =  v_{eff}[\Psi](\vect{r}) -  v_{eff}[\{\phi_i^{KS} \}](\vect{r})  -v_H[\rho](\vect{r}) , \label{rks_lps2}
\end{equation}
which is the equation (expressed in natural orbitals)  for the exchange-correlation potential used in the  modified RKS method (mRKS) \cite{Ospadov_2017} and it is the same as Eq. (\ref{vxc_psi}). Now, since the mRKS method uses only the LPS potential $ v_{eff}[\Psi](\vect{r})$ and $ v_{eff}[\{\phi_i\}](\vect{r})$ so the resulting exchange-correlation potential is  expected to be well behaved. This is indeed observed in application \cite{Ospadov_2017} of the mRKS method to the Ar atom .
\section{conclusion}
Previous work in the literature has shown that use of wavefunction based formula derived from LPS formulation leads to highly accurate exchange-correlation potential from wavefunctions calculated by expansion in finite basis set. In this study we have proved analytically and demonstrated numerically a general result: that the use of properly constructed approximate wavefunction - whether given in a functional form or in terms of basis-set expansion - in the LPS expression for potential leads to good approximation to the exact exchange-correlation potential  for  a given hamiltonian specified by $v_{ext}(\vect{r})$. Furthermore, we have shown that the difference between the exchange-correlation potential so obtained and that calculated by the inversion of the corresponding approximate density arises from the difference between  $v_{ext}(\vect{r})$ and the potential $\overline{v}_{ext}(\vect{r})$ corresponding to a given density. Our work thus extends the previous studies to all kinds of approximate wavefunctions and it gives a method to calculate accurate exchange-correlation potential by employing these. Additionally, we have also shown that the use of the LPS effective potential obtained from approximate wavefunction in the corresponding equation gives a density which is more accurate than that given by the wavefunction itself. This may pave the way to calculating accurate densities by employing approximate wavefunctions.
\setcounter{equation}{0}
\renewcommand\theequation{A.\arabic{equation}}
\section*{Appendix: LPS potential calculated from Slater determinant wavefunction}
\label{apx}
In this section we calculate LPS effective potential for the $N$ particle Slater-determinant wavefunction
\begin{equation}
\begin{split}
\Phi_{S,N}(\vect{x},\vect{x}_{2-N})
= \frac{1}{\sqrt{N!}}
\begin{bmatrix}
\phi_{1}(\vect{x}) & \phi_{2}(\vect{x}) & \dots  & \phi_{N}(\vect{x})  \\
\phi_{1}(\vect{x}_2) & \phi_{2}(\vect{x}_2) & \dots  & \phi_{N}(\vect{x}_2)  \\
 \vdots & \vdots  & & \vdots \\
\phi_{1}(\vect{x}_N) & \phi_{2}(\vect{x}_N) & \dots  & \phi_{N}(\vect{x}_N)
\end{bmatrix}
\end{split}
\end{equation}
constructed using one particle  orthogonal spin-orbitals $\{\phi_{i}(\vect{x})\}$. For the Hartree-Fock spin-orbitals $\{\phi_{i}(\vect{x}) = \phi^{HF}_{i}(\vect{x}) \}$ those are solution of  Hartree-Fock (HF) equation
\begin{equation}
\begin{split}
\Big[-\frac{\nabla^2}{2} +v_{ext}(\vect{r}) &+ \int \frac{\rho(\vect{r}')}{|\vect{r}-\vect{r}'|}d\vect{r}'  \\
&+\hat{v}^{HF}_{X} (\vect{x},\vect{x}') \Big]\phi^{HF}_i(\vect{x}) = \epsilon_i ^{HF}\phi^{HF}_i(\vect{x}), \label{hf_eqn}
\end{split}
\end{equation}
with $ \epsilon_i ^{HF}$ being the egienenergy corresponding to $\phi^{HF}_i$. $\Phi_{S,N}$ is an approximation to ground state wavefunction for interacting system and it is known as HF Slater determinant wavefunction. Here $\hat{v}^{HF}_{X} (\vect{x},\vect{x}')$ is HF exchange operator and it operates on spin-orbital $\phi^{HF}_i(\vect{x})$ as
\begin{align}
\hat{v}^{HF}_{X} &(\vect{x},\vect{x}')\phi^{HF}_i(\vect{x})\notag \\ &= -\sum_j  \int \frac{\phi^{HF}_j(\vect{x}) \phi^{*HF}_j(\vect{x}')\phi^{HF}_i(\vect{x}')}{|\vect{r}-\vect{r}'|}d\vect{x}'.
\end{align}
Similarly  if one employs $\{\phi_{i}(\vect{x}) =\phi^{KS}_{i}(\vect{x})\}$ with $\{\phi^{KS}_{i}(\vect{x})\}$ being solution of the Kohn-Sham  equation then $\Phi_{S,N}$ represents the ground-state wavefunction of the corresponding Kohn-Sham system. 
\par  For the calculation purpose we consider LPS potential in reduced density-matrix representation. In reduced density-matrix representation the  $P^{th}$ order reduced density-matrix $\gamma_P(\vect{x}',\vect{x}'_{2-P};\vect{x},\vect{x}_{2-P})$ is   defined using manybody wavefunction $\Psi$ as \cite{Yang}
\begin{equation}
\begin{split}
\gamma_P&(\vect{x}',\vect{x}'_{2-P};\vect{x},\vect{x}_{2-P} )= \\
&\frac{N!}{P!(N-P)!} \int \Psi(\vect{x}',\vect{x}'_{2-P},\vect{x}_{P+1-N}) \Psi^*(\vect{x},\vect{x}_{2-N}) d\vect{x}_{P+1-N}. 
\end{split}
\end{equation}
In particular for  $\Psi =\Phi_{S,N} $, the  $P^{th}$ order reduced density-matrix $\gamma_P(\vect{x}',\vect{x}'_{2-P};\vect{x},\vect{x}_{2-P})$ is related to the first order reduced density matrix  $\gamma_1(\vect{x}';\vect{x})$ by
\begin{equation}
\begin{split}
\gamma_P(\vect{x}',\vect{x}'_{2-P};\vect{x},\vect{x}_{2-P})= \hspace{6cm} \\
\frac{1}{\sqrt{P!}}
\begin{bmatrix}
\gamma_1(\vect{x}';\vect{x}) &\gamma_1(\vect{x}';\vect{x}_2) & \dots  & \gamma_1(\vect{x}';\vect{x}_P)  \\
\gamma_1(\vect{x}'_2;\vect{x}) &\gamma_1(\vect{x}_2';\vect{x}_2) & \dots  & \gamma_1(\vect{x}'_2;\vect{x}_P) \\
                                     \vdots & \vdots  & & \vdots \\
\gamma_1(\vect{x}'_P;\vect{x}) &\gamma_1(\vect{x}_P';\vect{x}_2) & \dots  & \gamma_1(\vect{x}'_P;\vect{x}_P) 
\end{bmatrix},  \hspace{1cm} 
\end{split}
\end{equation}
where $\gamma_1(\vect{x}';\vect{x}) = \sum_{i=1}^{N} \phi_{i}(\vect{x}')\phi^*_{i}(\vect{x})$. The density $\rho(\vect{r})$ in reduced density representation is calculated by
\begin{align}
\rho(\vect{r}) = \int \gamma_1(\vect{x};\vect{x}) d\sigma.
\end{align} 
Using the above relations one finds that different term of  LPS  potential in Eq.(\ref{LPS_mb})for Slater determinant wavefunction $\Phi_{S,N}$  are 
 \begin{equation}
 \begin{split}
\int \frac{\rho_{N-1}(\vect{r};\vect{r}')}{|\vect{r}-\vect{r}'|} d\vect{r}'& = \frac{2}{\rho(\vect{r})} \int\frac{\gamma_2(\vect{x},\vect{x}';\vect{x},\vect{x}')}{|\vect{r}-\vect{r}'|} d\vect{x}' d\sigma  \\ &=v_H(\vect{r}) +v_S(\vect{r}), \label{lps_first}
\end{split}
\end{equation}
\begin{widetext}
\begin{equation}
\begin{split}
\langle& \Phi_{N-1}|H_{N-1} -E^0_{N-1}|\Phi_{N-1}\rangle  \\
&= \frac{1}{\rho(\vect{r})}\Big\{2\int \Big[-\frac{\nabla'^2}{2} +v_{ext}(\vect{r}') \Big]\gamma_2(\vect{x},\vect{x}';\vect{x},\vect{x}'') \Big |_{\vect{x}' =\vect{x}''}d\vect{x}' d\sigma+3\int\frac{\gamma_3(\vect{x},\vect{x}',\vect{x}'';\vect{x},\vect{x}',\vect{x}'')}{|\vect{r}'-\vect{r}''|} d\vect{x}'d\vect{x}'' d\sigma\Big\}  -E^0_{N-1} \\ \\
&=  \Big\{\int \Big[-\frac{\nabla'^2}{2} +v_{ext}(\vect{r}') \Big]\gamma_1(\vect{x}';\vect{x}'') \Big |_{\vect{x}' =\vect{x}''}d\vect{x}' 
\\&  \hspace{2cm}+\frac{1}{2}\int\frac{\Big[\gamma_1(\vect{x}';\vect{x}')\gamma_1(\vect{x}'';\vect{x}'')-\gamma_1(\vect{x}';\vect{x}'')\gamma_1(\vect{x}'';\vect{x}')\Big]}{|\vect{r}'-\vect{r}''|} d\vect{x}'d\vect{x}'' -E^0_{N-1}\Big\}    \\
&\hspace{2cm}-\frac{1}{\rho(\vect{r})} \Big\{\int \Big[-\frac{\nabla'^2}{2} +v_{ext}(\vect{r}') \Big]\gamma_1(\vect{x}';\vect{x}) \gamma_1(\vect{x};\vect{x}'')\Big |_{\vect{x}' =\vect{x}''}d\vect{x}' d\sigma
  \\
&\hspace{2cm}+\int\frac{\gamma_1(\vect{x}';\vect{x})\Big[\gamma_1(\vect{x}'';\vect{x}'')\gamma_1(\vect{x};\vect{x}')-\gamma_1(\vect{x};\vect{x}'')\gamma_1(\vect{x}'';\vect{x}')\Big]}{|\vect{r}'-\vect{r}''|} d\vect{x}'d\vect{x}''d\sigma \Big\}    \\ \\
&=  \Big\{\sum_{i=1}^{N}\int \phi_i^*(\vect{x}') \Big[-\frac{\nabla'^2}{2} +v_{ext}(\vect{r}') \Big] \phi_i(\vect{x}')d\vect{x}' 
+\frac{1}{2}\int\frac{\rho(\vect{r}')\rho(\vect{r}'')}{|\vect{r}'-\vect{r}''|} d\vect{r}'d\vect{r}'' \\
& \hspace{2cm} -\frac{1}{2}\sum_{i=1,j=1}^{N}\int\frac{\phi_i^*(\vect{x}'')\phi_i^*(\vect{x}')\phi_j^*(\vect{x}')\phi_j^*(\vect{x}'')}{|\vect{r}'-\vect{r}''|} d\vect{x}'d\vect{x}'' -E^0_{N-1}\Big\}  \\
& \hspace{2cm}-\sum_{i=1,j=1}^{N}\int \frac{\phi^*_i(\vect{x})\phi_j(\vect{x})}{\rho(\vect{r})}d\sigma \Big\{\int \phi^*_j(\vect{x}')\Big[-\frac{\nabla'^2}{2} +v_{ext}(\vect{r}') +\int\frac{\rho(\vect{r}'')}{|\vect{r}'-\vect{r}''|}d\vect{r}'' \Big]\phi_i(\vect{x}') d\vect{x}'
\\
& \hspace{2cm}-\sum_{k=1}^{N}\int\frac{\phi^*_j(\vect{x}'')\phi_k(\vect{x}'')\phi^*_k(\vect{x}')\phi_i(\vect{x}')}{|\vect{r}'-\vect{r}''|} d\vect{x}'d\vect{x}'' \Big\}\label{lps_second}
\end{split}
\end{equation}
and 
\begin{equation}
\frac{1}{2}\int |\nabla \Phi_{N-1} |^2d\sigma d\vect{x}_2...d\vect{x}_{N} 
= \frac{1}{2\rho(\vect{r})}\int \nabla \nabla'\gamma_1(\vect{x};\vect{x}')\Big |_{\vect{x} =\vect{x}'}d\sigma - \frac{1}{8} \Big | \frac{\nabla \rho(\vect{r})}{\rho(\vect{r})}\Big|^2  =\frac{1}{2 \rho (\vect{r})}\int \sum_{i=1}^{N}  |\nabla\phi_i(\vect{x})|^2 d\sigma - \frac{1}{8} \Big | \frac{\nabla \rho(\vect{r})}{\rho(\vect{r})}\Big|^2.   \label{lps_third}
\end{equation}
\end{widetext}
In Eq. (\ref{lps_first}) $v_S(\vect{r})$ is known as Slater potential \cite{Slater_1951} and it is given by
\begin{equation}
\begin{split}
v_S(\vect{r}) &= - \frac{1}{\rho(\vect{r})} \int \frac{\gamma_1(\vect{x}';\vect{x}) \gamma_1(\vect{x};\vect{x}')}{|\vect{r}-\vect{r}'|}d\vect{x}'d\sigma  \\
&= \frac{1}{\rho(\vect{r})}\sum_{i=1,j=1}^{N} \int \frac{\phi^*_i(\vect{x})\phi_i(\vect{x}')\phi^*_j(\vect{x}')\phi_j(\vect{x})}{|\vect{r}-\vect{r}'|}d\vect{x}'d\sigma 
\end{split}
\end{equation} 
Having calculated different quantities of LPS potential in Eq. (\ref{LPS_mb}), on the applying  Eqs. (\ref{lps_first},\ref{lps_second},\ref{lps_third})  with Eq. (\ref{hf_eqn}) for Hartree-Fock wavefunction $\{\phi(\vect{x})= \phi^{HF}(\vect{x}) \}$ the  corresponding LPS potential is found to be
\begin{equation}
\begin{split}
v_{eff}^{HF}(\vect{r})& = v_H(\vect{r})+v_S(\vect{r})\\
&+\frac{1}{\rho^{HF} (\vect{r})}\int \sum_{i=1}^{N}(\mu^{HF}-\epsilon_i^{HF})|\phi^{HF}_i(\vect{x})|^2 d\sigma  \\
&+\frac{1}{2 \rho^{HF} (\vect{r})}\int \sum_{i=1}^{N}  |\nabla\phi_i^{HF}(\vect{x})|^2 d\sigma \\ 
& - \frac{1}{8} \Big | \frac{\nabla \rho^{HF}(\vect{r})}{\rho^{HF}(\vect{r})}\Big|^2
\label{lps_HF}
\end{split}
\end{equation}
The quantity $\mu^{HF}$ is calculated using 
\begin{align}
\mu^{HF} = E_{N}^{0,HF}-E_{N-1}^{0,HF},
\end{align} 
is the chemical potential of HF system in Koopman's approximation. 
In the calculation of LPS potential for the Kohn-Sham system $\{\phi(\vect{x})= \phi^{KS}(\vect{x}) \}$ and all the terms corresponding to interaction term $\frac{1}{|\vect{r}-\vect{r}'|}$ drop out of Eq. (\ref{lps_HF}). Then using the
Eqs. (\ref{lps_second},\ref{lps_third}) leads to
\begin{equation}
\begin{split}
 v_{eff}^{Pauli}(\vect{r})& = \frac{1}{\rho^{KS} (\vect{r})}\int \sum_{i=1}^{N}(\mu^{KS}-\epsilon_i^{KS})|\phi^{KS}_i(\vect{x})|^2 d\sigma  \\
&+\frac{1}{2 \rho^{KS} (\vect{r})}\int \sum_{i=1}^{N}  |\nabla\phi_i^{KS}(\vect{x})|^2 d\sigma - \frac{1}{8} \Big | \frac{\nabla \rho^{KS}(\vect{r})}{\rho^{KS}(\vect{r})}\Big|^2.
\end{split}
\end{equation}
Here $\mu ^{KS} = \epsilon_{max}^{KS}$ is eigenenergy of highest occupied Kohn-Sham orbital.
\bibliography{citations,akmybib,shorttitles} 
\end{document}